\documentclass{IEEEcsmag}

\usepackage[colorlinks,urlcolor=blue,linkcolor=blue,citecolor=blue]{hyperref}
\usepackage[T1]{fontenc} %
\usepackage{upmath}
\usepackage{booktabs}
\usepackage{tabularx}
\PassOptionsToPackage{hyphens}{url}\usepackage{hyperref}
\newcommand{\totalusecases}{89~}

\newcommand{\scientificusecases}[1]{6}
\newcommand{\characteristics}{24~}
\newcommand{\majorrev}[1]{\textcolor{black}{#1}}
\usepackage{setspace}

\jvol{XX}
\jnum{XX}
\paper{8}
\jmonth{September}
\jname{IEEE Software}
\pubyear{2020}

\begin{document}

\title{Serverless Applications:\\
        Why, When, and How?}

\author{Simon Eismann}
\affil{Julius-Maximilian University}

\author{Joel Scheuner}
\affil{Chalmers | University of Gothenburg}

\author{Erwin van Eyk}
\affil{Vrije Universiteit}

\author{Maximilian Schwinger}
\affil{German Aerospace Center}

\author{Johannes Grohmann}
\affil{Julius-Maximilian University}

\author{Nikolas Herbst}
\affil{Julius-Maximilian University}

\author{Cristina L. Abad}
\affil{Escuela Superior Politecnica del Litoral}

\author{Alexandru Iosup}
\affil{Vrije Universiteit}

\begin{abstract} %
Serverless computing shows good promise for efficiency and ease-of-use. Yet, there are only a few, scattered and sometimes conflicting reports on questions such as {\it Why do so many companies adopt serverless?}, {\it When are serverless applications well suited?}, and {\it How are serverless applications currently implemented?} To address these questions, we analyze 89 serverless applications from open-source projects, industrial sources, academic literature, and scientific computing---the most extensive study to date.
\end{abstract}

\maketitle

\begin{figure*}[!htb]
    \centering
    \includegraphics[width=\linewidth]{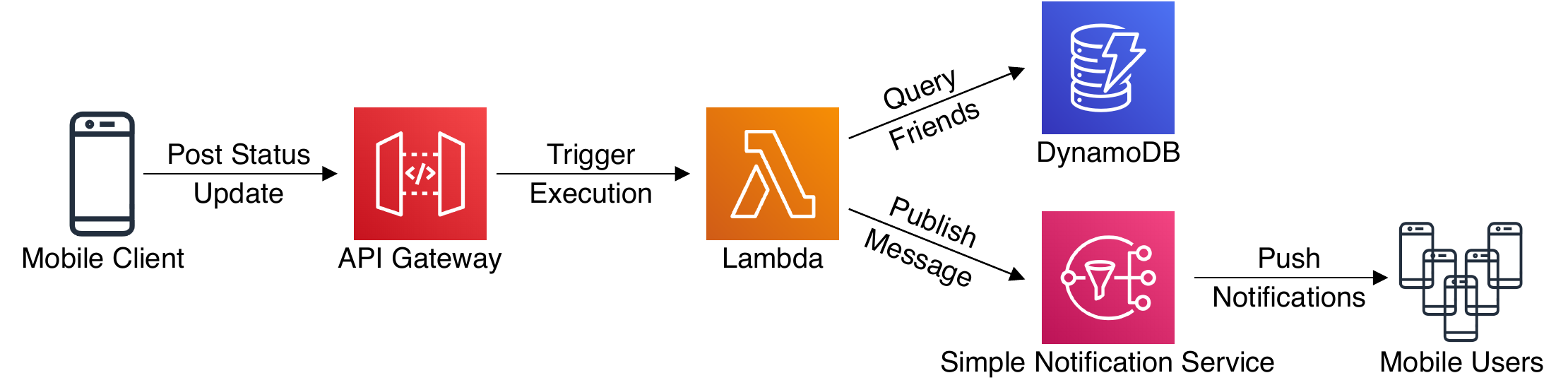}
    \caption{\majorrev{\textbf{[SIDEBAR, page 3] Example serverless application: mobile backend for a social media app.}
    In this example application, 
    a social media user wants to publish a status update, which should be seen by all the user's friends. 
    In 2019, this happened over a billion times a day on social media platforms such as Facebook, Twitter, and Instagram.
    Technology-wise, this would happen through a four-step process: (1) the user would compose the status update using the mobile clients of the social media platform, then (2) the user would send the status update using the mobile client, (3) the platform would orchestrate the operations needed to propagate the update inside the social media platform and to the user's friends, (4) each friend would receive the update on their social media clients, for example, on their mobile phones. Step 3 is the ``secret sauce'' of the social media platform -- although the users and their friends never see it, the software, and the resources (the {\it servers}) on which the software run, ensure the technical sustainability of the social media platform.
    \newline \newline
    With {\it conventional technology}, for step 3 the platform operators would need not only to develop the logic of routing the status update to each of the user's friends, but also to carefully manage the resources (the servers) on which the logic (the {\it software}) can run and to make sure the logic runs correctly. Resource management, and in particular resource provisioning and allocation, is a long-lasting hard problem in computing. Obstacles such as scaling the resources proportionally to the number of users (and their friends) are exacerbated by the fine-grained nature of each operation. Running the logic is also challenging, when subject to strict performance requirements---after all, the user expects all their friends to see the status update immediately, and to reply to it in a matter of seconds.
    \newline \newline
    With {\it serverless technology}, the cloud provider abstracts away the server management, provisioning servers with fine granularity, on demand, and with a {\it pay-per-use} model.
    To benefit from this, the serverless software gets transformed.
    Simplified, the process of posting updates proceeds as follows.
    The users post status updates using their mobile clients as a HTTP request to the API Gateway~(which operates as {\it serverless request routing}, that is, a specialized function that only routes requests and runs very efficiently without further developer intervention). Step 3 proceeds: the API Gateway %
    triggers 
    a lambda function ({\it serverless compute}), which queries the user's friends from a DynamoDB table ({\it serverless storage}) and publishes the status update to friends using the Simple Notification Service~(SNS)~({\it serverless publish/subscribe}). Finally, SNS generates push notifications with the status update for the user's friends. 
    Tens of other serverless functions get invoked, to authenticate the user, to authorize posting messages, to copy data between various locations, etc. All of these are orchestrated %
    on resources managed by the cloud operator.
    The serverless functions are fine grained, which leads to higher scalability than the coarser conventional approaches, but at the cost of more complex orchestration.
    For further examples of serverless applications, we refer to our technical report~\cite{techreport}.}}
    \label{fig:example}
\end{figure*}

\majorrev{\chapterinitial{Ease-of-use and efficiency} are two of the most desirable properties of software services. Ease-of-use entices more people to try the services and allows more to continue using them. Efficiency allows increased and longer operation of the service and, as the scale of software services has already reached a significant fraction of the world's energy consumption, keeps these services sustainable.
But ease-of-use and efficiency have been historically at odds with each other: The easier the software is to use, the fewer hints it can provide to the platform on which it runs, leaving it to solve complex problems of resource (in particular, of server) management. Although not a universal remedy, serverless computing aims to provide both ease-of-use and efficiency for common software services that run on cloud resources. We investigate in this article why?, when?, and how are serverless applications useful?}

\majorrev{Serverless computing is any computing platform that \textit{hides server usage from developers and runs code on-demand automatically scaled and billed only for the time the code is running~}\cite{Castro2019}.  
Sidebar~\ref{fig:example} exemplifies how serverless can be used to implement the mobile backend for a social media app. 
More generally, serverless applications combine managed stateless ephemeral compute solutions such as AWS Lambda, Azure Functions, or Google Cloud Functions~(Function-as-a-Service, FaaS), and fully provider-managed services for messaging, file storage, databases, streaming, or authentication~(Backend-as-a-Service, BaaS).} Serverless computing is increasingly adopted by industry~\cite{Market2017,Leitner2019} and studied by academics~\cite{Eyk2019,cidr}.
One crucial reason is that serverless operations empower developers to focus on implementing business logic and letting the cloud providers handle all operational concerns, such as deployment, resource allocation, and autoscaling~\cite{Castro2019}.

However, only a few, and sometimes conflicting, reports address important questions such as \textit{Why are practitioners choosing to build serverless applications?}, \textit{When are serverless applications well suited?}, or \textit{How are serverless applications implemented in practice?}. \majorrev{For example, there are reports of significant cost savings by switching to serverless applications~\cite{Adzic2017, Levinson2020}, but also articles suggesting higher cost in some scenarios compared to traditional hosting~\cite{Eivy2017}.} Having concrete information on these topics would be valuable for managers to guide decisions on whether a serverless application can be a suitable solution for a specific use case. The SPEC-RG Cloud group suggests that surveys of real-world serverless computing are needed to understand architectural, implementation, and deployment patterns emerging in the ``serverless soup''~\cite{Eyk2019}. Leitner et al. also discuss that empirical studies about serverless use are required to guide software developers with building serverless solutions~\cite{Leitner2019}.

\begin{figure*}[t]
    \centering
    \includegraphics[width=0.9\linewidth]{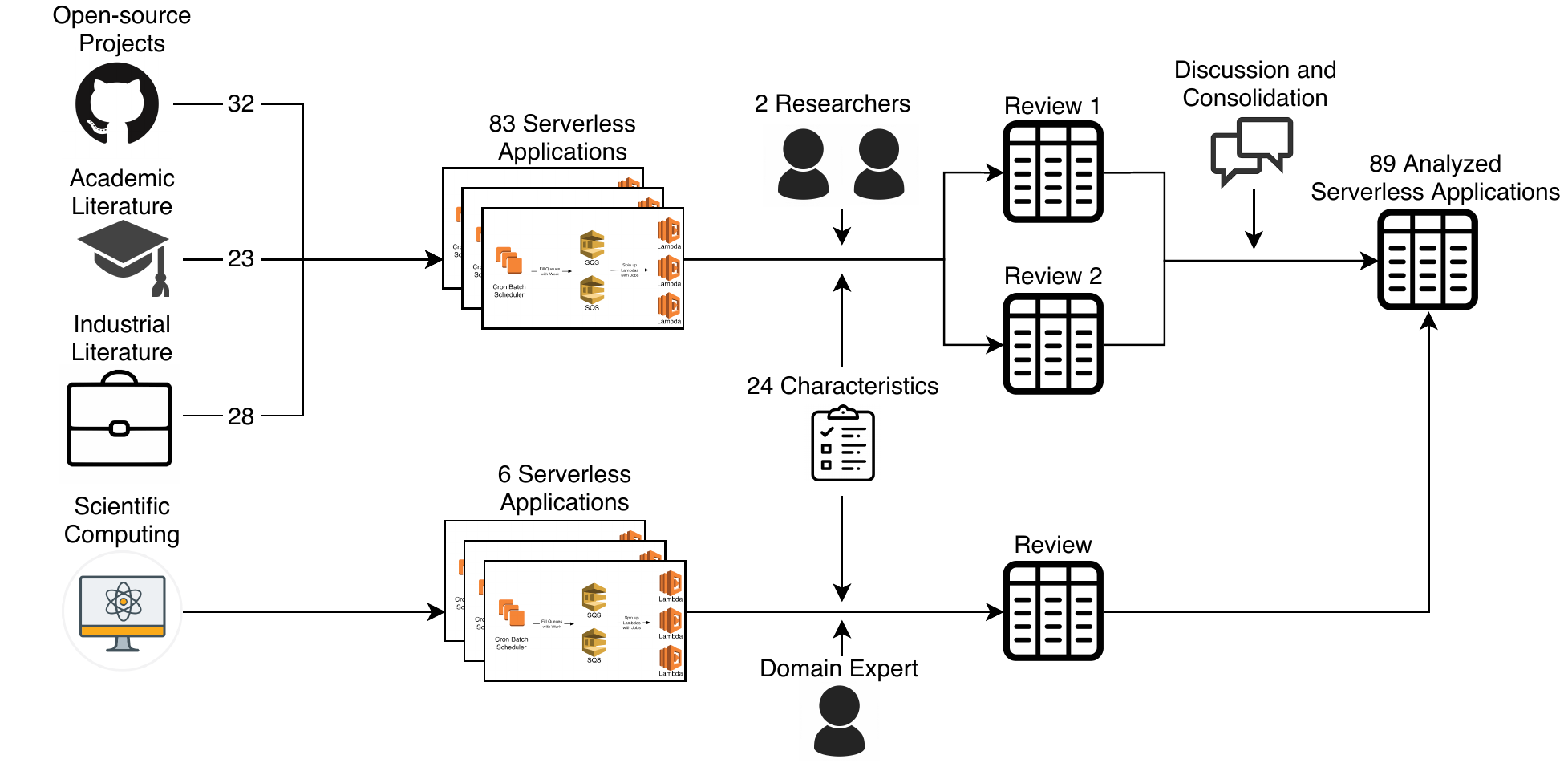}
    \caption{\textbf{[SIDEBAR] Methodology for serverless application collection and characterization.} We collected descriptions of  serverless application from four sources: open-source projects, academic literature, industrial literature, and scientific computing. Next, we randomly assigned two out of the seven total reviewers to review each serverless application based on a set of fixed characteristics. In the following discussion and consolidation phase, we discussed and resolved any differences between the two resulting characteristics reviews. For the scientific applications, a different approach was necessary, as many of them were not publicly available yet. Therefore, these applications are reviewed by a single domain expert, which is either involved in the development of the applications or in direct contact with the development team. If the information to determine a characteristic for a serverless application was not available, we labeled the characteristic as ''Unknown'' for this application. \majorrev{The percentage of ''Unknowns'' ranges from 0–19\%  with two outliers at 25\% and 30\% for the characteristics presented in this article.} These ''Unknowns'' were excluded for the percentage values presented in this article. For a more detailed breakdown of our results and a more in-depth description of our methodology, we refer to our technical report~\cite{techreport}.}
    \label{fig:overview}
\end{figure*}

Addressing the need for a comprehensive empirical study of serverless applications, we collect a total of \totalusecases descriptions of existing serverless applications. We focus on diverse sources, e.g., GitHub projects, blog posts, scientific publications, or talks at industry conferences. We analyze every serverless application regarding \characteristics characteristics (see Sidebar~\ref{fig:overview} for a description of our methodology).
Based on this data, we investigate why are practitioners choosing to adopt the serverless paradigm, when are serverless applications used, and how are they implemented in practice. \majorrev{The percentage values  for the characteristics presented in the remainder of this article exclude any serverless application for which we could not determine the respective characteristic.}

    \label{tbl:use cases}

\section{Why are practitioners choosing to adopt serverless?}
\noindent Pioneers and journalists seem to agree on several potential benefits of serverless applications: reduced operation effort, faster development due to the heavy use of Backend-as-a-Service, and near-infinite scalability of serverless applications. Many also discuss significant cost savings from switching to serverless. However, not all these benefits hold in general. \majorrev{For example, cost savings have come under scrutiny~\cite{Eivy2017}.}
To understand why practitioners choose to adopt serverless, we investigate the descriptions and documentation of applications in our dataset.

For 27 of the applications in our dataset, we cannot discern the motivation for going serverless. 
Of the remaining 62 applications, 47\% choose serverless to save costs. Serverless holds the promise to save costs due to its pay-per-use model for irregular or bursty workloads, which would have low resource utilization and thus higher cost with traditional hosting options. This concurs with our observation that 84\% of the serverless applications have bursty workloads.

Another reason for serverless adoption, which applies for 34\% of the applications in our dataset, is that developers no longer need to bother with operational concerns, such as deployment, scaling, or monitoring, and instead can focus on developing new features. \majorrev{This was reported not only for utility functionality, such as a CI/CD pipeline or malware detection, where traditionally servers need to be maintained for only few and irregular executions. Interestingly, it was also reported for user-facing APIs serving large-scale traffic, where one would expect other concerns such as performance or availability to take priority.}

\majorrev{A third reason (also 34\%) is the scalability of serverless applications.}
Although traditional applications can also be scalable, serverless applications offer near-infinite, out-of-the-box scalability with minimal engineering effort. This is because the FaaS implementation commonly used for serverless computing is fine-grained and thus conveniently %
parallel.

\majorrev{Based on our dataset, the most commonly reported reasons for the adoption of serverless are to save costs for irregular or bursty workloads, to avoid operational concerns, and for built-in scalability.} Other reasons, such as improved performance and faster time-to-market, are less common (19\% and 13\%, respectively).

\majorrev{These findings are mostly in accordance with a recent community survey where over 160 participants~\cite{Levinson2020} report that the positive impacts of adopting a serverless architecture are the adoption of an event-driven architecture~(51\%), cost of resources~(44\%), speed of development~(36\%), flexibility of scaling~(31\%) and application performance~(19\%).}

\section{When are serverless applications used?}
\noindent When starting a project, managers need to decide which technology stack is best suited. Currently, managers are faced with the difficult question if serverless is well suited for a specific application. A common assumption is that serverless applications are best suited for utility functionality and less applicable for latency-critical, high-volume core functionality. For example, Netflix uses AWS Lambda for utility functionality, such as video encoding, file backup, security audits of EC2 instances, and monitoring. However, the core functionality such as the website/app backend or video delivery, is still running on traditional IaaS cloud services~\cite{netflix}.

Contrary to this popular belief, we found that serverless applications are not limited to utility functionality. Our data indicates that many serverless applications implement utility functionality (39\%), but also that many serverless applications implement core functionality~(42\%) and scientific workloads~(16\%). This is consistent with our finding that a substantial number of serverless applications~(39\%) experience high traffic intensity, a high proportion complemented by on-demand applications experiencing a low traffic intensity~(47\%), and by scheduled applications~(17\%). %

A common argument against serverless applications is that cold starts make them unsuitable for applications with latency requirements. 
To the contrary, we find that serverless applications are used for latency critical tasks, despite the cold starts affecting tail latencies.
Concretely, %
38\% of the surveyed serverless applications have no latency requirements. However, 32\% of the serverless applications have latency requirements for all functionality, 28\% have partial latency requirements, and 2\% even have real-time requirements. 

Another argument is that current serverless platforms can be unsuited for long-running tasks or tasks with large data volumes~\cite{cidr}. Our dataset supports this hypothesis as 69\% of the surveyed serverless applications have a data volume of less than 10 MB, and 75\% have an execution time in the range of seconds. To overcome this limitation, the area needs further innovation.

To summarize, serverless applications are most commonly used for short-running tasks with low data volume and bursty workload. However, contrary to popular belief, serverless applications are also widely used for latency-critical, high-volume core functionality. Overall, there are examples of serverless applications across all application types, requirements, and workloads.

\begin{figure}[t!]
    \centering
    \includegraphics[width=1\linewidth]{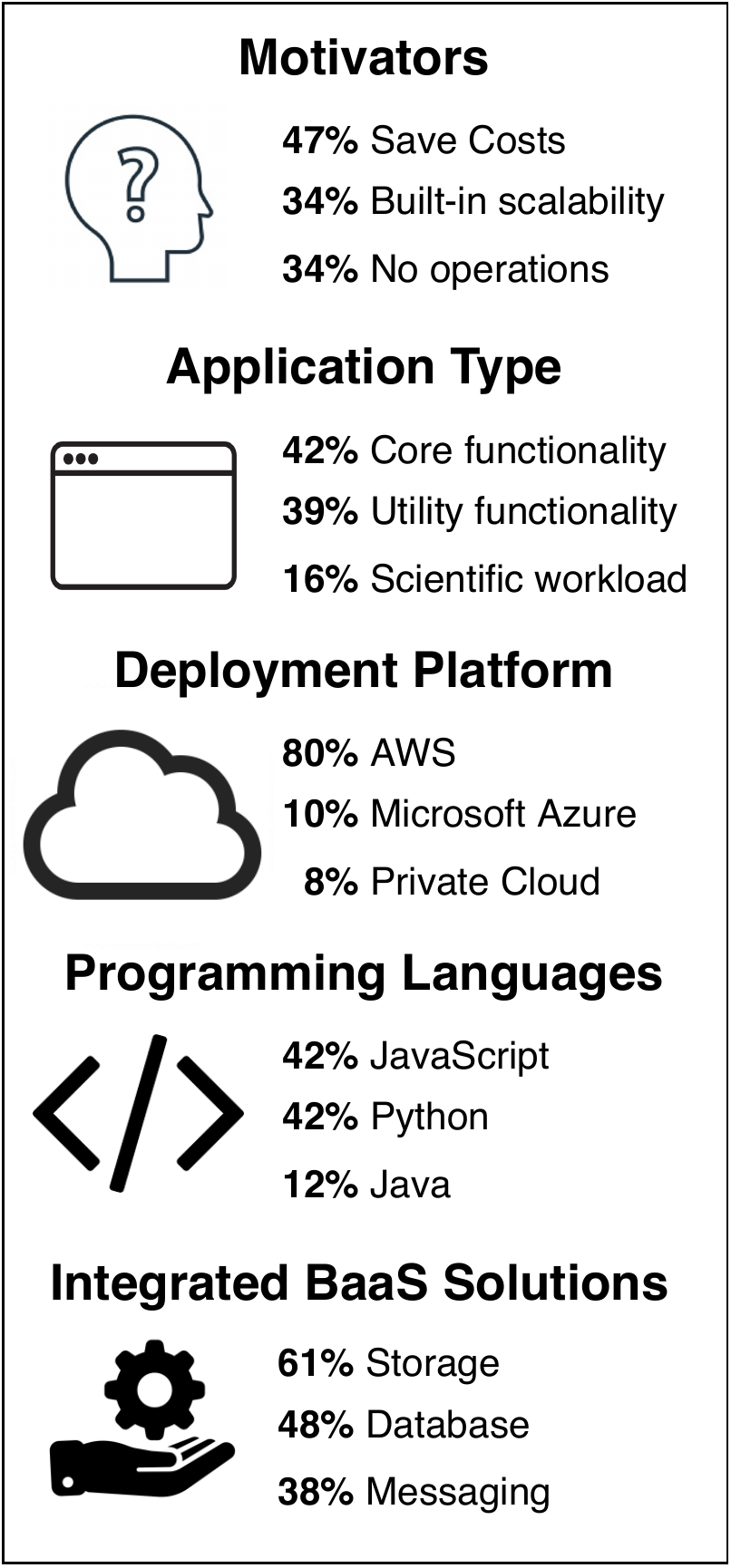}
    \caption{\textbf{Key findings.} The results limited to the top 3 values and a single application can have multiple values for motivators, programming languages, and integrated backend-as-a-service solutions. For more detailed results, we refer to our technical report~\cite{techreport}.}
    \label{fig:results}
\end{figure}

\section{How are serverless applications implemented in practice?}
\noindent When implementing a serverless application, engineers face several technology and architecture decisions, such as selecting the cloud platform, serverless platform, programming language, Backend-as-a-Service options, and appropriate granularity level for serverless functions.
Based on our collection of serverless applications, we identify the most popular approaches to implementing serverless applications.

Among the surveyed applications, AWS is by far the most popular deployment platform chosen by 80\% of the applications.
The other cloud vendors are less represented, with Azure at 10\%, IBM at 7\%, and Google Cloud at 3\%. We see two potential reasons for this vast lead: 
(a) AWS introduced their FaaS platform (AWS Lambda) two years before the other cloud vendors, so their platform is likely to be the most mature, and 
(b) AWS has the largest market share of general cloud computing~\cite{gartner}, which gives it a larger existing user base that can move applications to serverless. About 8\% of the surveyed serverless applications run in private clouds; however, these are mostly applications from academia and scientific computing. The low adoption of private clouds is in stark contrast to the large number of existing open-source Function-as-a-Service frameworks~\cite{Eyk2019}. Part of the appeal of the serverless application model is the automation of operational concerns. We hypothesize that the increase in such concerns that comes with maintaining a fleet of servers and an open-source Function-as-a-Service framework is deterring the adoption of these frameworks. Additionally, most serverless applications make use of managed services (storage, databases, messaging, logging, streaming, etc.), which are not available directly in a private cloud environment. 

Serverless platforms support popular programming languages. \majorrev{In our study, we were able to determine the programming language for 67 of the 89 serverless application.} JavaScript~(42\%) and Python~(42\%) were by far the most popular programming languages. Some applications are also written in Java~(12\%), C/C++~(11\%), or C\#~(8\%), while only few use Go~(5\%) or Ruby~(2\%). Traditionally, interpreted languages such as JavaScript and Python have lower cold-start times, that is, time required to initialize a new instance, than compiled languages. \majorrev{However, as the technology matures, this difference seems to level out~\cite{coldstarts} and there are multiple new approaches for cold start mitigation. For example, AWS introduced the reserved capacity feature, which provides prewarmed function instances to avoid cold starts. The adoption of such cold start mitigation strategies could lead to an increased usage of compiled languages.}

Cloud providers offer managed services as part of BaaS, such as messaging, file storage, databases, streaming, logging, and machine learning. In our study, we found that the most popular external services are cloud storage~(e.g., S3) and databases~(e.g., Dynamo DB). As serverless functions are ephemeral and stateless, they need to rely on external services to persist data and manage state. Many serverless applications also use some form of managed messaging~(38\%), such as managed pub/sub (17\%), streaming~(11\%), or queues~(10\%). Messaging services are popular, because serverless functions rarely communicate via external calls, as this results in double billing~\cite{trilemma}. The popularity of messaging services is consistent with our finding that serverless applications mostly use event-based architectures~(60\%) or workflow engines~(38\%) to coordinate multiple functions. It is interesting to note that only 12\% of serverless applications use no BaaS solutions, highlighting the symbiosis between FaaS and BaaS. 

The appropriate granularity of serverless functions is currently debated~\cite{faasification}. Opinions range from wrapping each program function as a serverless function, or each API endpoint as a serverless function, to the full-scale conversion of each microservices into serverless functions. We find that 82\% of the surveyed serverless applications consist of five functions or less and that 93\% consist of ten functions or less. With one exception, the remaining applications consist of less than 20 functions. \majorrev{Therefore, the granularity of a serverless function is more akin to a full microservice or an API endpoint in our dataset. Consequently, the term ``serverless function'' might be somewhat of a misnomer as they are not related to the general programming concept of functions.}

To summarize, serverless applications are mostly implemented on AWS, in either Python or JavaScript, and commonly make use of cloud storage, managed databases, and messaging services. \majorrev{Also, serverless functions are not comparable to programming functions in terms of size according to our dataset.} \newline

\majorrev{\chapterinitial{B}ased on our analysis of 89 serverless applications, we find that the most commonly reported reasons for the adoption of serverless are to save costs for irregular or bursty workloads, to avoid operational concerns, and for the built-in scalability.} Serverless applications are most commonly used for short-running tasks with low data volume and bursty workloads but are also frequently used for latency-critical, high-volume core functionality. Serverless applications are mostly implemented on AWS, in either Python or JavaScript, 
and make heavy use of BaaS.
We see this study as a step towards a community-wide policy of sharing and discussion of serverless applications.
Such a catalog of serverless applications could stimulate a new wave of serverless designs, facilitate meaningful tuning and performance benchmarking, and overall prove useful for both industry and academia.

\begin{IEEEbiography}{Simon Eismann}{\,} is currently a PhD student at the chair of software engineering at the University of Würzburg. He received the M.S. degree from the University of Würzburg in 2017. His research interests include cloud computing, serverless and performance analysis/modeling. Contact him at simon.eismann@uni-wuerzburg.de.
\end{IEEEbiography}

\begin{IEEEbiography}{Joel Scheuner}{\,} is a PhD student at the joint division of software engineering at Chalmers University of Technology and the University of Gothenburg.
He received an M.S. in Software Systems from the University of Zurich in 2017.
His research interests include cloud computing, performance engineering, and software engineering.
Contact him at scheuner@chalmers.se.
\end{IEEEbiography}

\begin{IEEEbiography}{Erwin van Eyk}{\,}
is a PhD student at Vrije Universiteit Amsterdam, the Netherlands, and the chair of the SPEC-RG Cloud Serverless activity. In 2019, he received an M.Sc. degree from TU Delft, the Netherlands, for work on cloud computing and serverless workflows.
\end{IEEEbiography}
\begin{IEEEbiography}{Maximilian Schwinger}{\,} is a PhD student at the chair for software engineering at the University of Würzburg and received his Diploma in Computer Science from the TU Munich in 2006. Since then he is working for the German Aerospace Center (DLR) as a Software and Systems Engineer. His research interest includes high performance computing, cloud based computing, and scientific computing in the domain of satellite based earth observation. Contact him at maximilian.schwinger@dlr.de.\end{IEEEbiography}
\begin{IEEEbiography}{Johannes Grohmann}{\,} is currently a PhD student at the chair of software engineering at the University of Würzburg. He received the M.S. degree from the University of Würzburg in 2016. His research interests include serverless and cloud computing, and performance model learning and analysis. Contact him at johannes.grohmann@uni-wuerzburg.de.\end{IEEEbiography}
\begin{IEEEbiography}{Nikolas Herbst}{\,}is a research group leader at the chair of software engineering at the University of W\"urzburg. He received a PhD from the University of W\"urzburg in 2018 and serves as elected vice-chair of the SPEC Research Cloud Group. His research topics include predictive data analysis, elasticity in cloud computing, auto-scaling and resource management, performance evaluation of virtualized environments, autonomic and self-aware computing. Contact him at nikolas.herbst@uni-wuerzburg.de.
\end{IEEEbiography}
\begin{IEEEbiography}{%
Cristina L. Abad}{\,} is associate professor at Escuela Superior Politecnica del Litoral, ESPOL, in Ecuador, where she leads the Distributed Systems Research Lab (DiSEL). She obtained MS and PhD in CS degrees from the University of Illinois at Urbana-Champaign, funded in part through Fulbright and Computer Science Excellence Fellowships. At ESPOL, she has received two Google Faculty Research Awards. Her main research interests lie at the intersection of distributed systems and performance engineering. Contact her at cabad@fiec.espol.edu.ec. \end{IEEEbiography}
\begin{IEEEbiography}{%
Alexandru Iosup}{\,}
is full professor and University Research Chair at Vrije Universiteit Amsterdam, and member of the Young Royal Academy of Arts and Sciences of the Netherlands. 
He is the chair of the Massivizing Computer Systems research group at the VU and of the SPEC-RG Cloud group. His work in distributed systems and ecosystems has received prestigious recognition, including the 2016 Netherlands ICT Researcher of the Year, the 2015 Netherlands Higher-Education Teacher of the Year, and several SPEC community awards SPECtacular (last in 2017). He can be contacted at A.Iosup@vu.nl or @AIosup. 
\end{IEEEbiography}
\end{document}